\documentclass[fleqn,12pt,twoside]{article}
\usepackage{espcrc1}
\usepackage{graphicx}
\usepackage[figuresright]{rotating}
\newcommand{\AmS}{{\protect\the\textfont2
  A\kern-.1667em\lower.5ex\hbox{M}\kern-.125emS}}

\title{
\vspace{-2.5cm}
       {\normalsize \hfill ITEP-LAT/2002--29}   \\[-0.2cm]
       {\normalsize \hfill KANAZAWA 02--41}   \\[0.8cm]
Finite temperature phase transition in lattice QCD with $N_f=2$ nonperturbatively improved Wilson fermions at $N_t=8$
}

\author{Y. Mori\address[KITP]{{\footnotesize Institute for Theoretical Physics, Kanazawa University, Kanazawa 920-1192, Japan}},
V. Bornyakov\addressmark\thanks{On leave of absence from IHEP, Protvino,
Russia}, M. Chernodub\addressmark, Y. Koma\addressmark, Y.
Nakamura\addressmark, T. Suzuki\addressmark, 
M.~Polikarpov\address[ITEP]{{\footnotesize Institute of Theoretical and
Experimental Physics, B.Cheremushkinskaya 25, Moscow, 117259, Russia}} , D.
Sigaev\addressmark, P. Uvarov\addressmark, A. Veselov\addressmark, A.
Slavnov\address{{\footnotesize Steklov Mathematical Institute, Vavilova, 42,
117333 Moscow, Russia}}, G. Schierholz\address[DESY]{{\footnotesize NIC/DESY
Zeuthen, Platanenallee 6, 15738 Zeuthen, Germany and Deutsches
Elektronen-Synchrotron DESY D-22603 Hamburg, Germany}} and 
H.~St\"uben\address{{\footnotesize Konrad-Zuse-Zentrum f\"ur Informationstechnik
Berlin, D-14195 Berlin, Germany}} }

\begin{document}

% typeset front matter
\maketitle

\begin{abstract}
The finite temperature lattice QCD with $N_f=2$ nonperturbatively  improved Wilson fermions is studied on $16^3 8$ lattice.
Using abelian projection after fixing to MA gauge
we determine the transition temperature for  $m_{\pi}/m_{\rho} \sim 0.8$.
\end{abstract}

%=============================================================================
\section{Introduction}
\vspace{-0.1cm}
Determination of the critical temperature of the chiral phase transition 
is one of the important nonperturbative problems in QCD to be addressed
in lattice simulations.
Bielefeld group and CP-PACS collaborations using two different types of
improved lattice action for fermions were able to estimate $T_c$ in the
chiral limit and their values are in good agreement
\cite{Karsch:2000kv,AliKhan:2000iz}.
Still there are many sources of systematic uncertainties
and new computations of $T_c$  with different actions are useful as an
additional check.
We made first large scale simulations of the nonperturbatively $O(a)$ improved
Wilson fermion action.
Moreover we performed simulations with the
lattice spacing $a$ substantially smaller than in studies by Bielefeld and
CP-PACS groups.

The following fermionic action is  employed in our study:
%==================================
\begin{equation}
S_F = S^{(0)}_F - \frac{\rm i}{2} \kappa\, g\, c_{sw} a^5 \sum_x \bar{\psi}(x)\sigma_{\mu\nu}F_{\mu\nu}\psi(x),
\label{fermact}
\end{equation}
%==================================
where $S^{(0)}_F$ is the original Wilson action, $c_{sw}$ is the
clover coefficient  determined nonperturbatively \cite{Jansen:1998mx}.
%==================================
We use Wilson gauge field action.

The action (\ref{fermact}) has been used in $T=0$ studies of lattice QCD by
UKQCD and QCDSF collaborations \cite{Booth:2001qp,Allton:2002sk}. These studies
confirmed that $O(a)$ lattice artifacts are suppressed as expected. To fix the
physical scale and $m_{\pi}/m_{\rho}$ ratio we used results obtained by these
collaborations \cite{Booth:2001qp}. So far only $N_t=4$ and $6$ finite
temperature results obtained with action (\ref{fermact}) are available
\cite{Edwards:1999mm}. These results were obtained at a rather large quark mass
($m_{\pi}/m_{\rho}>0.85$). In this work me make simulations with $N_t=8$ what
allows us to decrease $m_{\pi}/m_{\rho}$ down to $\sim 0.8$. We choose the
spatial extension of the lattice $N_s=16$ as a compromise between computational
burden and need to reduce the finite size effects.

It is known that the order parameter of the finite temperature phase transition
in quenched QCD is the Polyakov loop and the corresponding symmetry is global
$Z(3)$ symmetry. In the chiral QCD the order parameter of the chiral symmetry
breaking transition is the chiral condensate $<\bar{\psi}\psi>$.   As numerical
results show \cite{Karsch:2000kv} both order parameters can be used to locate
the transition point at intermediate values of the quark mass.  We use Polyakov
loop susceptibility to determine the transition temperature.

\section{Simulation details}
Hybrid Monte Carlo algorithm with parameters
$\delta\tau=0.0125$, $ n_\tau=20$, providing acceptance rates of about 70\% is
used in our simulations. The simulations were done on SR8000 Hitachi at KEK,
Tsukuba and MVS 1000M at Joint Supercomputer Center, Moscow. For analysis SX5
NEC at RCNP and PC-cluster at ITP, Kanazawa were employed. Our code performs at
the speed of  2.4 GFlops  per node on Hitachi computer. We needed from 1000
($\tau \equiv N_{traj} \cdot n_\tau \cdot \delta\tau =250$) to 3000 
($\tau=750$) trajectories for
thermalization, depending on $\kappa$ and $\beta$. For runs started from
configurations generated at the adjacent $\kappa$ this value was much lower.
We determined the transition temperature at two values of $\beta$, 5.2 and
5.25, varying $\kappa$ as shown in Table 1.

\begin{table}[tb]
\vspace{-0.5cm}
\begin{center}
\newcommand{\m}{\hphantom{$-$}}
\newcommand{\cc}[1]{\multicolumn{1}{c}{#1}}
\renewcommand{\tabcolsep}{0.4pc} % enlarge column spacing
\renewcommand{\arraystretch}{0.8} % enlarge line spacing
\begin{tabular}{c|c|c|c}
\hline
\multicolumn{2}{c|}{{\footnotesize $\beta=5.2, N_s=16$}} & \multicolumn{2}{|c}{{\footnotesize $\beta=5.25, N_s=16$}}  \\ \hline
 {\footnotesize $\kappa$} & {\footnotesize \# of traj.} & {\footnotesize $\kappa$} & {\footnotesize \# of traj.} \\ \hline
 {\footnotesize 0.1330}   & {\footnotesize 3409}  & {\footnotesize 0.1330 }  & {\footnotesize 1540 }   \\
 {\footnotesize 0.1335}   & {\footnotesize 4500}  & {\footnotesize 0.1335 }  & {\footnotesize 7439 }   \\
 {\footnotesize 0.1340}   & {\footnotesize 2100}  & {\footnotesize 0.13375}  & {\footnotesize 9225 }   \\
 {\footnotesize 0.1343}   & {\footnotesize 6650}   & {\footnotesize 0.1339 }  & {\footnotesize 12470}  \\
 {\footnotesize 0.1344}   & {\footnotesize 7485}  & {\footnotesize 0.1340 }  & {\footnotesize 19479}  \\
 {\footnotesize 0.1345}   & {\footnotesize 4647}    & {\footnotesize 0.1341 }  & {\footnotesize 13750}   \\
 {\footnotesize 0.1348}   & {\footnotesize 6013}    & {\footnotesize 0.13425}  & {\footnotesize 5155 }   \\
 {\footnotesize 0.1355}   & {\footnotesize 5650}    & {\footnotesize 0.1345 }  & {\footnotesize 2650 }   \\
 {\footnotesize 0.1360}   & {\footnotesize 3699}    & {\footnotesize 0.1350 }  & {\footnotesize 1780 }   \\
 \hline
\end{tabular}
\\
\vspace{0.2cm}
{\footnotesize Table.1~~Simulation statistics.}
\label{table:1}
\end{center}
\vspace{-1.0cm}
\end{table}

We fixed the maximally abelian (MA)  gauge on generated configurations. The
simulated annealing algorithm has been used to fix the gauge. The advantages of
this algorithm in comparison with the usual iterative algorithm has been
demonstrated in $SU(2)$ pure gauge theory for MA gauge \cite{Bali:1996dm} and
Maximal Center gauge \cite{Bornyakov:2000ig}.

Abelian and monopole Polyakov loops and their susceptibilities were measured on
gauge fixed configurations. We found that for abelian and monopole observables
the signal/noise ratio was better than that for gauge invariant nonabelian
observables. This observation is in agreement with the results from quenched
QCD at $T>0$ and both quenched and unquenched QCD at $T=0$. In particular the
maximum of the Polyakov loop susceptibility is essentially better separated
from the rest of the data for the monopole Polyakov loop than for nonabelian
Polyakov loop. It is important that the
transition temperature values determined by both susceptibilities are the same.

\begin{figure}[t]
\hspace{-0.0cm}
\begin{minipage}[t]{160mm}
\begin{center}
\begin{minipage}[t]{60mm}
\vspace{-4.75cm}
\includegraphics[width=5cm, height=4.5cm]{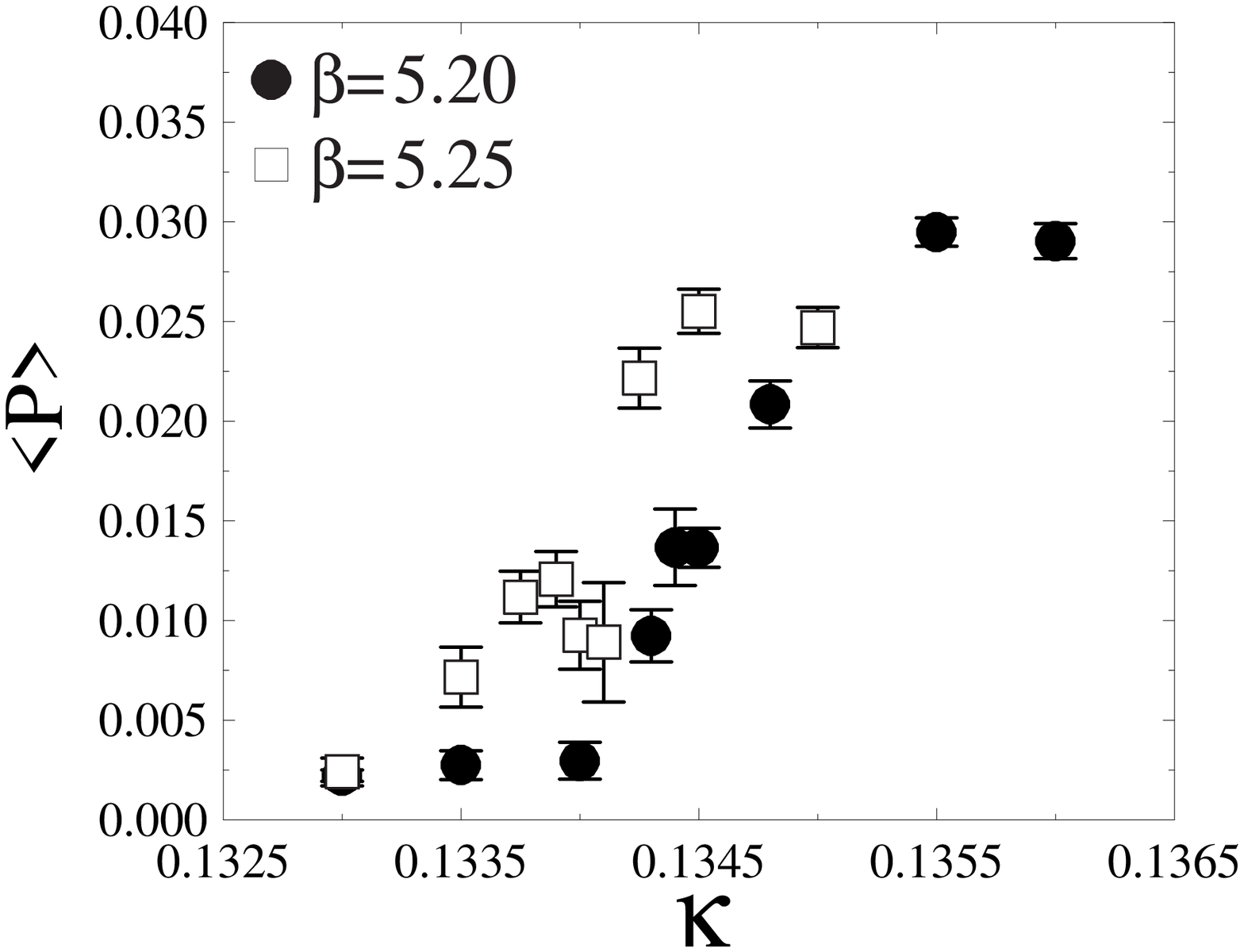}
\end{minipage}
\begin{minipage}[t]{60mm}
\hspace{1.0cm}
\includegraphics[width=5cm, height=4.8cm]{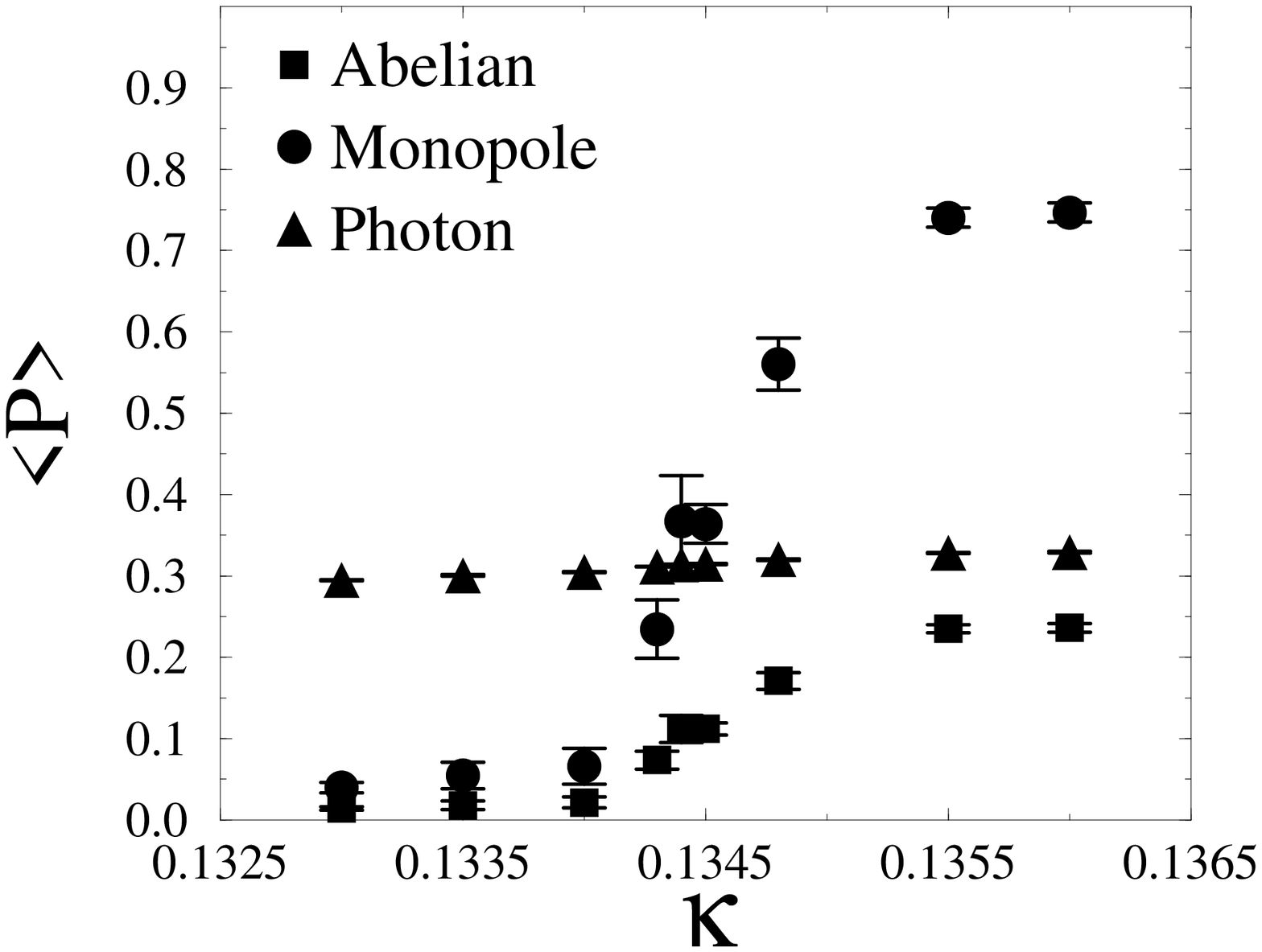}
\end{minipage}
\vspace{-1.0cm}
\caption{{\footnotesize Nonabelian Polyakov loop at both $\beta$'s (left), and
the abelian, monopole and photon Polyakov loops at $\beta=5.2$ (right).}}
\label{ploop}
\end{center}
\end{minipage}
\end{figure}

\section{Results and Conclusions}
\vspace{-0.1cm}
In Fig.\ref{ploop} we show results for average of various kinds of Polyakov loops.
One can see that $\langle P\rangle $ is a smooth function of $\kappa$
and that abelian and monopole Polyakov loops behavior is qualitatively
the same as behavior of the nonabelian Polyakov loop while the photon
Polyakov loop is almost constant across the transition.

\begin{figure}[b]
\hspace{-0.5cm}
\begin{center}
\begin{minipage}[t]{160mm}
\begin{center}
\includegraphics[width=5.0cm, height=5cm]{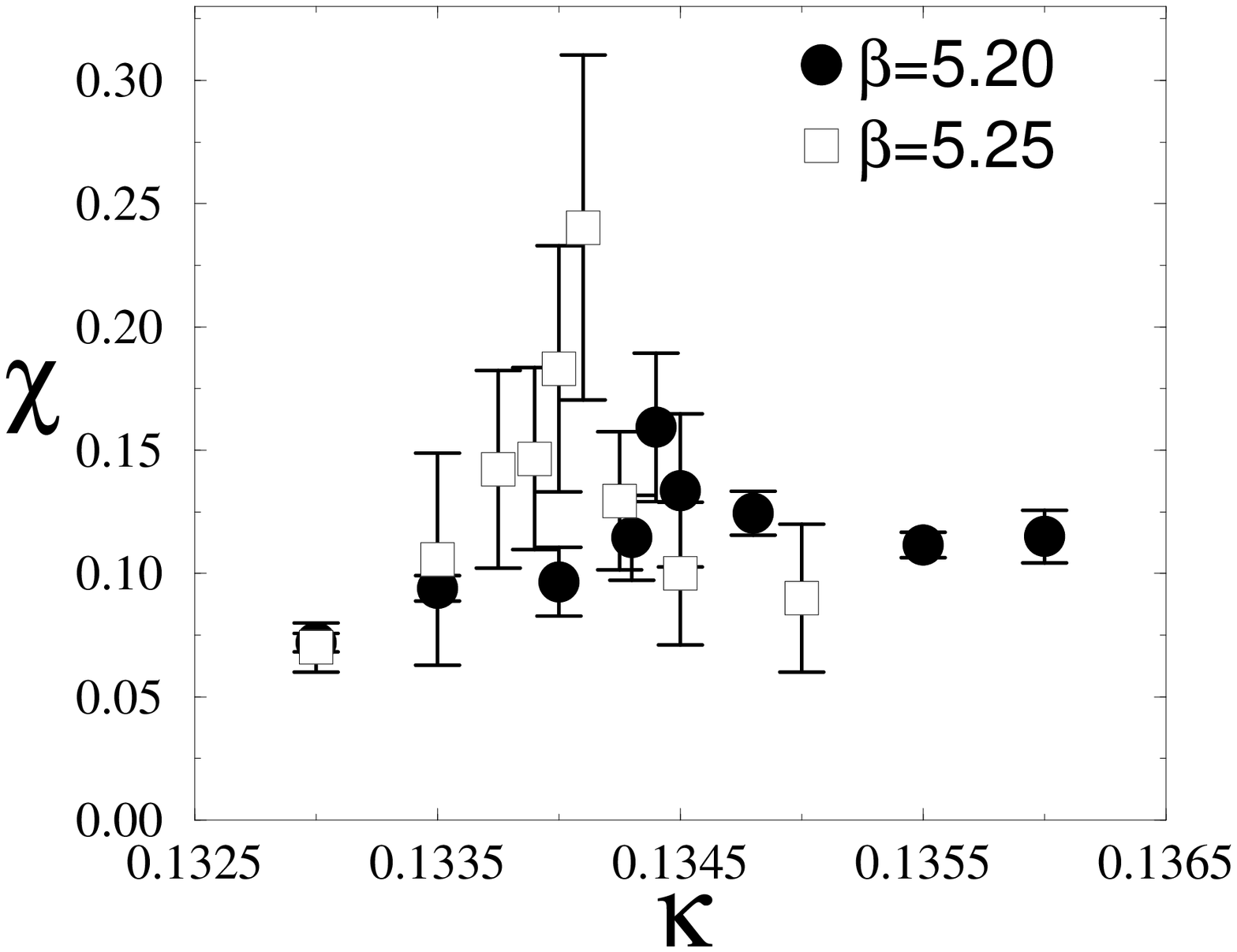}
\hspace{2.0cm}
\includegraphics[width=5.0cm, height=5cm]{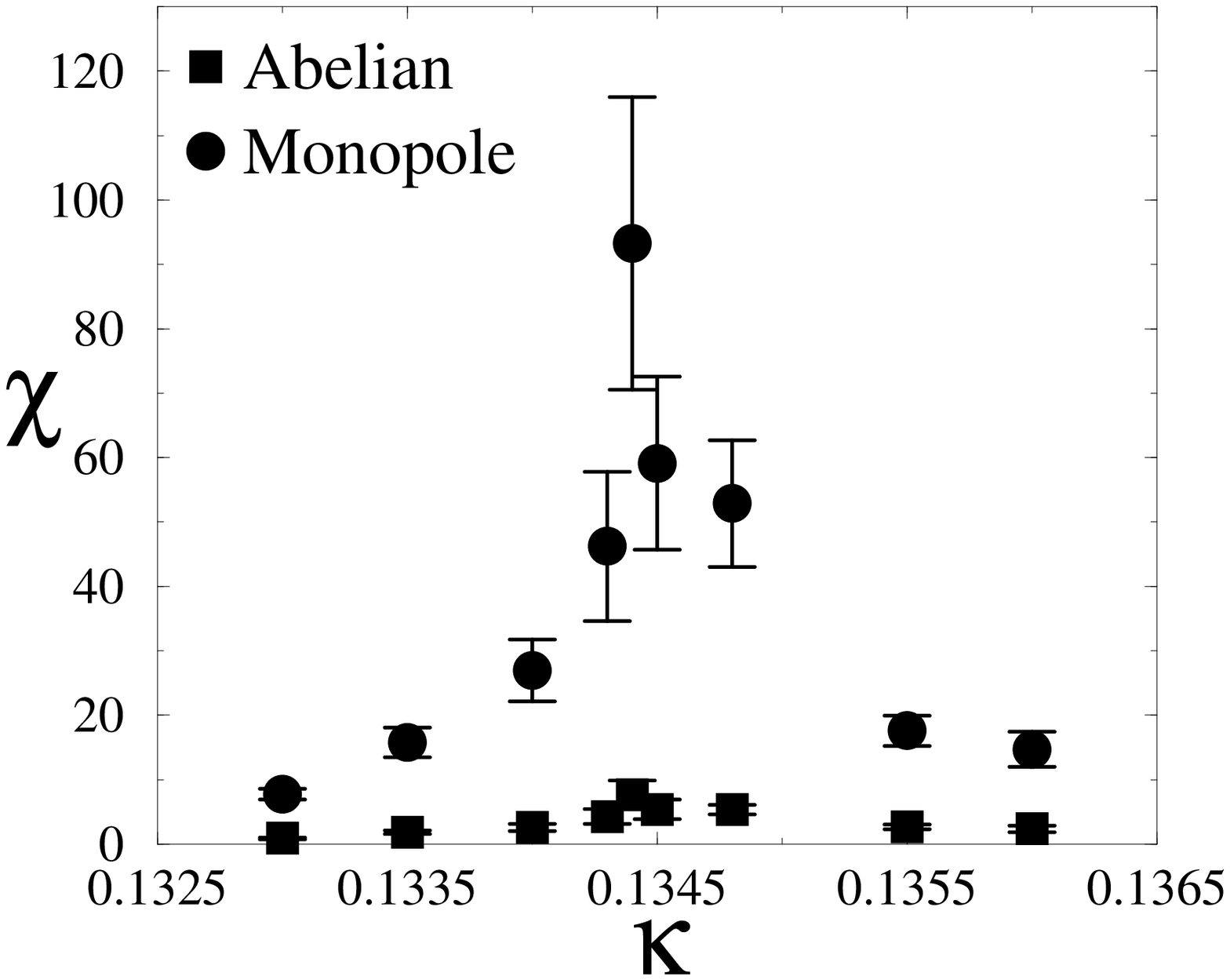}
\vspace{-1.4cm}
\hspace{-0.5cm}
\caption{{\footnotesize Nonabelian Polyakov loop susceptibilities at both $\beta$'s (left), abelian and monopole Polyakov loop susceptibilities at $\beta=5.2$
(right)}}
\label{susc1}
\end{center}
\end{minipage}
\end{center}
\end{figure}

Transition point $\kappa$, denoted as $\kappa_t$, has been determined from the
maximum of the Polyakov loop susceptibility, see Fig \ref{susc1}. As can be
seen from Fig.\ref{susc1} the susceptibilities for abelian and monopole
Polyakov loops have maxima at the same value of $\kappa$. We found that
$\kappa_t=0.1344(1)$ at $\beta=5.2$ and $\kappa_t=0.1341(1)$ at $\beta=5.25$,
see Fig.\ref{susc1}. This was transformed into transition temperature with the
help of the interpolation formula for $ r_0/a $ \cite{Booth:2001qp}. The
results are: $T_c r_0 = 0.54(2)$ and $0.56(2)$. In physical units, taking
$1/r_0=394 $ MeV, we obtained $T_c=213(10)$ and $222(10)$MeV, respectively.   
Using again data from \cite{Booth:2001qp} we estimate $m_{\pi}/m_{\rho}=0.78,
0.82$ at the transition points.

\begin{figure}[t]
\vspace{-1.0cm}
\begin{center}
\begin{minipage}[t]{60mm}
\hspace{0.0cm}
\includegraphics[width=5.2cm, height=4.8cm]{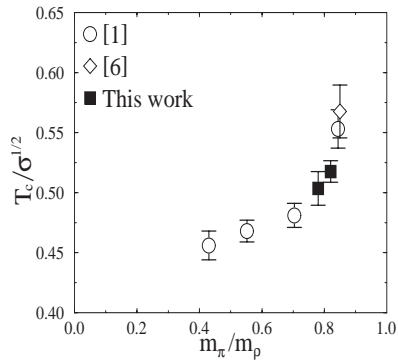}
\vspace{-1.0cm}
\caption{{\footnotesize Transition temperature}}
\label{TR}
\end{minipage}
\end{center}
\end{figure}
%\end{minipage}

In Fig.\ref{TR} we show our results for transition temperature in comparison
with results of refs. \cite{Karsch:2000kv} and \cite{Edwards:1999mm}. The ratio
$T_c/\sqrt{\sigma}$ was computed for $\sqrt{\sigma}=425$ MeV used in
\cite{Karsch:2000kv}. We conclude that our results are in good qualitative
agreement with previous results \cite{Karsch:2000kv}. 
This agreement implies that
the dependence of the transition temperature on the lattice spacing is rather
weak.

For our lattice with $N_s/N_t=2$ the question of the finite volume effects is
very important. To check this effect we made simulations at $\beta=5.2,
\kappa=0.1343$ on $N_s=24$ lattice. The $\kappa$ value was chosen close to the
transition point where finite volume effects should be more pronounced. We
found that both average of the Polyakov loop and its susceptibility are the
same within error bars as in our main simulations on $N_s=16$ lattice. This
implies that the finite volume effects should not spoil our conclusions.
We are planning to perform simulations on $24^3\cdot 10$ lattice at
$m_\pi/m_\rho < 0.7$ to determine $T_c$ closer to the chiral limit.\\

\noindent
{\bf Acknowledgements} \\
\vspace{-0.1cm}
\noindent
This work is partially
supported by grants INTAS-00-00111, RFBR 02-02-17308, RFBR
01-02-117456, RFBR 00-15-96-786 and CRDF award RPI-2364-MO-02.
We are very obliged to  the staff of the Joint Supercomputer
Center at Moscow and especially to A.V. Zabrodin for the help in
computations. We thank Ph. De Forcrand,
V. Mitrjushkin, M. M\"uller-Preussker for useful discussions.
M.Ch. is supported by JSPS Fellowship grant No. P01023. V.B. acknowledges
support by JSPS during period of work on this project.


\begin{thebibliography}{99}
\baselineskip 4.5mm
%\cite{Karsch:2000kv}
\bibitem{Karsch:2000kv}
{\footnotesize F.~Karsch, E.~Laermann and A.~Peikert,
%``Quark mass and flavor dependence of the QCD phase transition,''
Nucl.\ Phys.\ B {\bf 605} (2001) 579
[arXiv:hep-lat/0012023].}
%%CITATION = HEP-LAT 0012023;%%
%%%%%%%%%%%%%%%%%%%%%%%%%%%%%%%%%%%%%%%%%%%%%
%\cite{AliKhan:2000iz}
\bibitem{AliKhan:2000iz}
{\footnotesize A.~Ali Khan {\it et al.}  [CP-PACS Collaboration],
%``Phase structure and critical temperature of two flavor QCD with  renormalization group improved gauge action and clover improved Wilson  quark
action,''
Phys.\ Rev.\ D {\bf 63} (2001) 034502
[arXiv:hep-lat/0008011].}
%%CITATION = HEP-LAT 0008011;%%
%%%%%%%%%%%%%%%%%%%%%%%%%%%%%%%%%%%%%%%%%%%%%
%%CITATION = HEP-LAT 0103023;%%
%\cite{Jansen:1998mx}
\bibitem{Jansen:1998mx}
{\footnotesize K.~Jansen and R.~Sommer  [ALPHA collaboration],
%``O(alpha) improvement of lattice QCD with two flavors of Wilson quarks,''
Nucl.\ Phys.\ B {\bf 530} (1998) 185
[arXiv:hep-lat/9803017].}
%%CITATION = HEP-LAT 9803017;%%
%%%%%%%%%%%%%%%%%%%%%%%%%%%%%%%%%%%%%%%%%%%%%
%\cite{Booth:2001qp}
\bibitem{Booth:2001qp}
{\footnotesize S.~Booth {\it et al.}  [QCDSF-UKQCD collaboration],
%``Determination of $\Lambda_\overline{MS}$ from quenched and $N_f = 2$
%dynamical  QCD,''
Phys.\ Lett.\ B {\bf 519} (2001) 229
[arXiv:hep-lat/0103023].}
%%%%%%%%%%%%%%%%%%%%%%%%%%%%%%%%%%%%%%%%%%%%%
%\cite{Allton:2002sk}
\bibitem{Allton:2002sk}
{\footnotesize C.~R.~Allton {\it et al.}  [UKQCD Collaboration],
%``Effects of non-perturbatively improved dynamical fermions in QCD at
%fixed lattice spacing,''
Phys.\ Rev.\ D {\bf 65} (2002) 054502
[arXiv:hep-lat/0107021].}
%%CITATION = HEP-LAT 0107021;%%
%%%%%%%%%%%%%%%%%%%%%%%%%%%%%%%%%%%%%%%%%%%%%
%\cite{Edwards:1999mm}
\bibitem{Edwards:1999mm}
{\footnotesize R.~G.~Edwards and U.~M.~Heller,
%``Thermodynamics with dynamical clover fermions,''
Phys.\ Lett.\ B {\bf 462} (1999) 132
[arXiv:hep-lat/9905008].}
%%CITATION = HEP-LAT 9905008;%%
%%%%%%%%%%%%%%%%%%%%%%%%%%%%%%%%%%%%%%%%%%%%%
%\cite{Kaczmarek:2000mm}
\bibitem{Kaczmarek:2000mm}
{\footnotesize O.~Kaczmarek, F.~Karsch, E.~Laermann and M.~Lutgemeier,
%``Heavy quark potentials in quenched QCD at high temperature,''
Phys.\ Rev.\ D {\bf 62} (2000) 034021
[arXiv:hep-lat/9908010].}
%%CITATION = HEP-LAT 9908010;%%
%%%%%%%%%%%%%%%%%%%%%%%%%%%%%%%%%%%%%%%%%%%%%
%\cite{Bali:1996dm}
\bibitem{Bali:1996dm}
{\footnotesize G.~S.~Bali, V.~Bornyakov, M.~Muller-Preussker and K.~Schilling,
%``Dual Superconductor Scenario of Confinement: A Systematic Study of Gribov Copy Effects,''
Phys.\ Rev.\ D {\bf 54} (1996) 2863
[arXiv:hep-lat/9603012].}
%%CITATION = HEP-LAT 9603012;%%
%%%%%%%%%%%%%%%%%%%%%%%%%%%%%%%%%%%%%%%%%%%%%
%\cite{Bornyakov:2000ig}
\bibitem{Bornyakov:2000ig}
{\footnotesize V.~G.~Bornyakov, D.~A.~Komarov and M.~I.~Polikarpov,
%``P-vortices and drama of Gribov copies,''
Phys.\ Lett.\ B {\bf 497} (2001) 151
[arXiv:hep-lat/0009035].}
%%CITATION = HEP-LAT 0009035;%%

\end{thebibliography}
\end{document}